\documentclass[journal=nalefd,manuscript=article]{achemso}

\usepackage{amssymb} 
\usepackage{amsmath} 
\usepackage[utf8]{inputenc} 
\usepackage[sort&compress,numbers]{natbib}
\pdfoutput=1 

\usepackage{siunitx}  
\usepackage{graphicx}
\usepackage[export]{adjustbox}
\usepackage{hyperref}
\usepackage{float}

\author{Niklas Müller}
\affiliation{Department of Physics, University of Regensburg, 93040 Regensburg, Germany}
\alsoaffiliation{Institute of Physics, University of Oldenburg, 26129 Oldenburg, Germany}
\author{Soufiane el Kabil}
\affiliation{Department of Physics, University of Regensburg, 93040 Regensburg, Germany}
\author{Gerrit Vosse}
\affiliation{Institute of Physics, University of Oldenburg, 26129 Oldenburg, Germany}
\author{Lina Hansen}
\affiliation{Institute of Physics, University of Oldenburg, 26129 Oldenburg, Germany}
\author{Christopher Rathje}
\affiliation{Institute of Physics, University of Oldenburg, 26129 Oldenburg, Germany}
\author{Sascha Schäfer}
\email{sascha.schaefer@ur.de}
\affiliation{Department of Physics and Regensburg Center for Ultrafast Nanoscopy, University of Regensburg, 93040 Regensburg, Germany}

\title{Spectrally resolved free electron-light coupling strength in a transition metal dichalcogenide}


\begin{document}

\begin{abstract}
Recent advancements in electron microscopy have introduced innovative techniques enabling the inelastic interaction of fast electrons with tightly confined and intense light fields. These techniques, commonly summarized under the term photon-induced near-field electron microscopy now offer unprecedented capabilities for a precise mapping of the characteristics of optical near-fields with remarkable spatial resolution but their spectral resolution were only scarcely investigated.  In this study, we employ a strongly chirped and temporally broadband light pulse to investigate the interaction between free electrons and light at the edge of a MoS$_2$ thin film. Our approach unveils the details of electron-light coupling, revealing a pronounced dependence of the coupling strength on both the position and photon energy. Employing numerical simulations of a simplified model system we identify these modulations to be caused by optical interferences between the incident and reflected field as well as an optical mode guided within the transition metal dichalcogenide film.     
\end{abstract}

\section{Introduction}

Electron microscopy excels in providing imaging with superb spatial resolution down to atomic length scales. Besides its capability to pinpoint the precise location of atoms, electron microsccopy is able to provide information about specific electronic material excitations and properties using electron energy loss spectroscopy (EELS) \cite{egerton2011electron}. The substantial progress in the spectral resolution of electron spectrometers and monochromators down to the meV-regime \cite{krivanek2014vibrational} has opened up a plethora of new directions in EELS now being possible. Recent examples include the mapping of surface phonon polaritons \cite{li2021three,li2021direct}, and plasmons \cite{bosman2007mapping,nelayah2007mapping,schaffer2009high} and even the bonding and electronic configuration of single atoms \cite{suenaga2010atom,ramasse2013probing}. While the EELS signal results from the interaction between the imaging electron and its self-induced field in the material, recently a new technique, called photon induced near-field electron microscopy (PINEM) \cite{barwick2009photon,garcia2010multiphoton,park2010photon}, emerged in ultrafast transmission electron microscopy (UTEM) \cite{zewail2010four}. In PINEM, an external light source is applied to illuminated the material. The probing electron couples to the near-field of the illuminated sample and absorbs or emits photons from or to the field \cite{di2019probing}. PINEM is perfectly suited to study the near-field distributions of plasmonic structures and has been applied, e.g. to map the near-field around silver nano-wires \cite{barwick2009photon}, nano-particles \cite{yurtsever2012direct}, nano-tips \cite{feist2015quantum} and gold nano-stars \cite{liebtrau2021spontaneous}. Furthermore, polarization-dependent coupling between far-field light and near-field modes were studied \cite{piazza2015simultaneous,lummen2016imaging} and even the chirality of nano-structures can be probed by applying circularly polarized light fields \cite{harvey2020probing}. Since, in comparison to EELS, PINEM is a stimulated process it can exhibit a comparably large cross section with a notable depletion of the original electron energy state \cite{feist2015quantum}. Using the high temporal resolution of UTEM in combination with PINEM also enables the spatiotemporal imaging of material excitations like 2D polaritons \cite{kurman2021spatiotemporal}. Another advantage of PINEM is its high spectral resolution which is governed by the externally applied optical field and not the energy resolution of the electron spectrometer or the monochromaticity of the probing electron beam \cite{de2008electron}. One example with a spectral resolution in the µeV range was recently demonstrated for characterizing photonic modes in a microscale ring resonator using frequency-shifted monochromatic light \cite{henke2021integrated}. Such an approach is particularly well suited for assessing closely spaced mode frequencies within a small spectral window. By using optical parametric amplifiers, the excitation wavelength can be tuned over a broad spectral range allowing to study the excitation energy dependency of plasmon resonances in nano wires \cite{pomarico2017mev} or the optical modes within a photonic cavity \cite{wang2020coherent} but not without losing resolution due to the lights spectral line width. Furthermore, tuning of the wavelength can lead to a variation of the beam pointing and the beam profile between different wavelengths and a change of the average optical power on the sample. These difficulties can be avoided by using a chirped multi-color light pulse \cite{kfir2020controlling} creating a quasi monochromatic excitation field in time that can be probed by comparably short electron pulses. Therefore, spectrally resolved PINEM should be perfectly suited to examine materials with complex optical properties such as transition metal dichalcogenides (TMDC) which until now have been mostly studied by optical spectroscopy \cite{mak2010atomically,wang2018colloquium}. EELS investigations of excitons in TMDC materials have been scarce due to the dense low-energy excitation spectrum of these materials which requires high spectral resolution \cite{tizei2015exciton,nerl2017probing}. First cathodoluminesence spectroscopy (CL) experiments on TMDC materials are reported \cite{taleb2022charting,taleb2023phase}, although in this case the impact of electronic excitations by secondary electrons becomes important, even resulting in photon bunching \cite{meuret2015photon,Fiedler_2023}.
          
Here, we demonstrate the broadband wavelength-resolved electron-light coupling efficiency mediated by the local optical field at a MoS$_2$ flake edge. By using a strongly chirped and elongated broadband light pulse we show that the coupling strength has a pronounced spectral dependency resulting from interferences between the incident and scattered field and optical modes within the flake. Scanning the electron beam over the flakes edge then reveals the spatially and spectrally resolved optical near-field distribution. The experimental findings are compared to numerical simulations of the expected coupling strength spatial and spectral dependency at a MoS$_2$ thin film edge.

\section{Experimental Setup}

\begin{figure}[H]
	\includegraphics[scale=1]{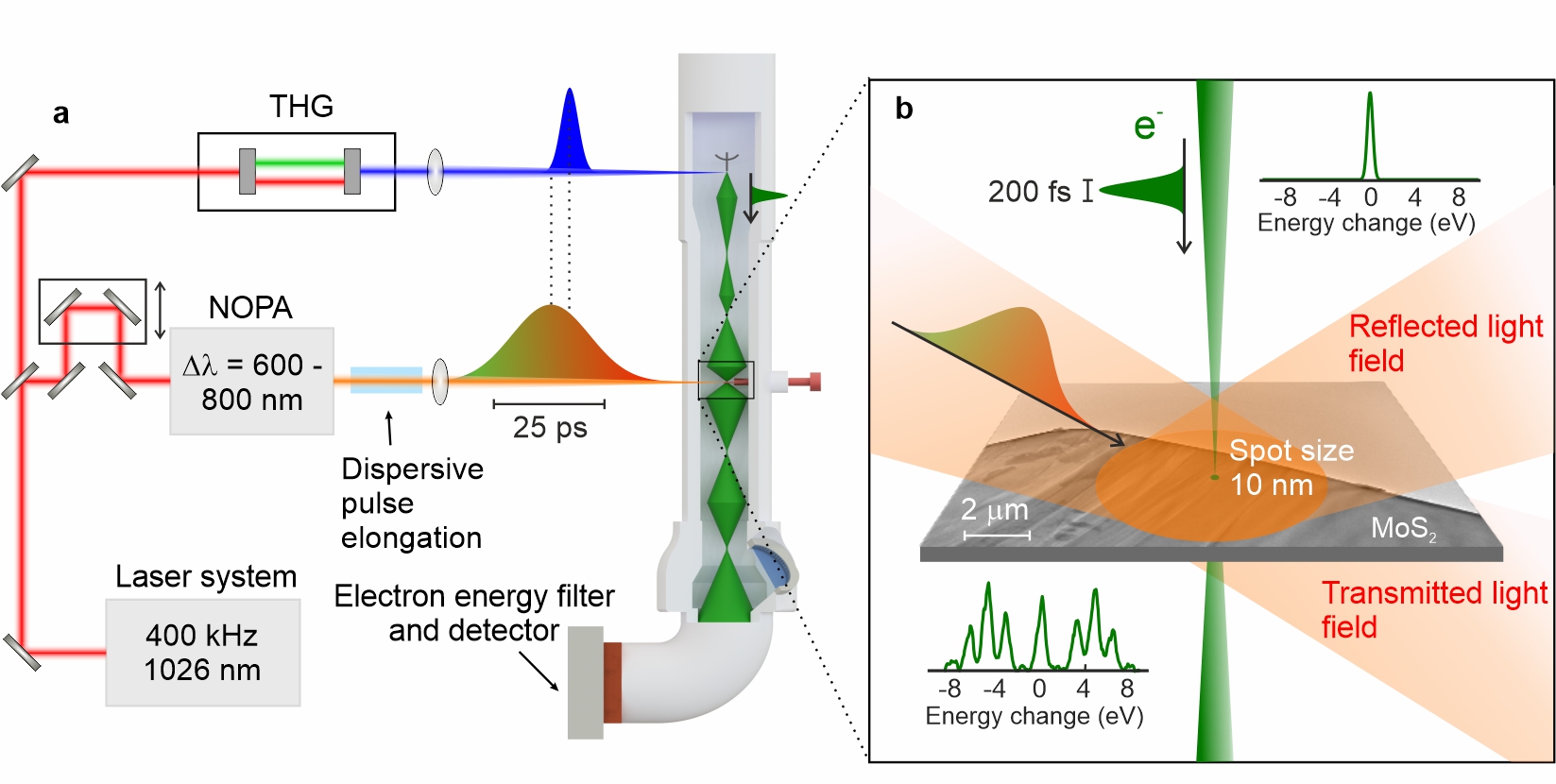}
	\caption{Experimental optical pump/ electron-probe setup and near-field-mediated electron-light interactions. \textbf{a} Schematic of the experimental setup consisting of an ultrafast transmission electron microscope in which femtosecond photoelectrons pulses are extracted from a cold-field emitter source illuminated by the third harmonic (THG: third harmonic generation) of a femtosecond pump laser. The strongly chirped broadband output of a non-collinear optical parametric amplifier (NOPA) is utilized to optically excited the sample at variable delay time. \textbf{b} The pulsed quasi-monochromatic electron beam passes through the optical near-field around the illuminated MoS$_2$ sample. Photon absorption and emission leads to sidebands in the electron energy spectrum with an energy spacing equal to the instantaneous photon energy. Experimental electron energy spectra before and after the near-field interaction are shown in the upper and lower inset, respectively. By changing the relative arrival time, the instantaneous photon energy of the optical field can be tuned allowing to probe the wavelength resolved electron-light coupling strength without changing the average optical power. }
	\label{figure1}
\end{figure}

For the nanoscale mapping of the local electron-light coupling strength at an illuminated MoS$_2$ edge, we utilize the inelastic scattering of femtosecond electron pulses (200-fs electron pulse width, 200-keV electron energy) at the edge near-field. The optical near-field is driven by a up-chirped broadband light pulse (25~ps pulse width, 150~nm spectral width (10$\%$), incidents angle $\alpha$~=~57$^{\circ}$) and probed by electron pulses which are derived from a laser-driven cold-field emitter and focused to diameters of 10~nm at the sample position (see Fig. 1a).  
An electron passing through the optical near-field can absorb and emit photons from the optical near-field resulting in photon sidebands in the electron energy spectrum after the interaction (Fig. 1b). For the current sample, optical spot diameters of about 50~µm and pulse energies of 8.75~nJ, up to about four sidebands are observed on the electron energy gain and loss side,respectively. Due to the large chirp of the optical pump pulse (group delay dispersion of about 54300~fs$^2$), the electron experiences a quasi-monochromatic near-field with the instantaneous photon energy depending on the delay between electron and light pulses. Thereby, without changing the average optical power on the sample, the electron-light coupling strength can be probed for different photon energies by adjusting this delay.

\begin{figure}[H]
	\center
	\includegraphics[scale=1]{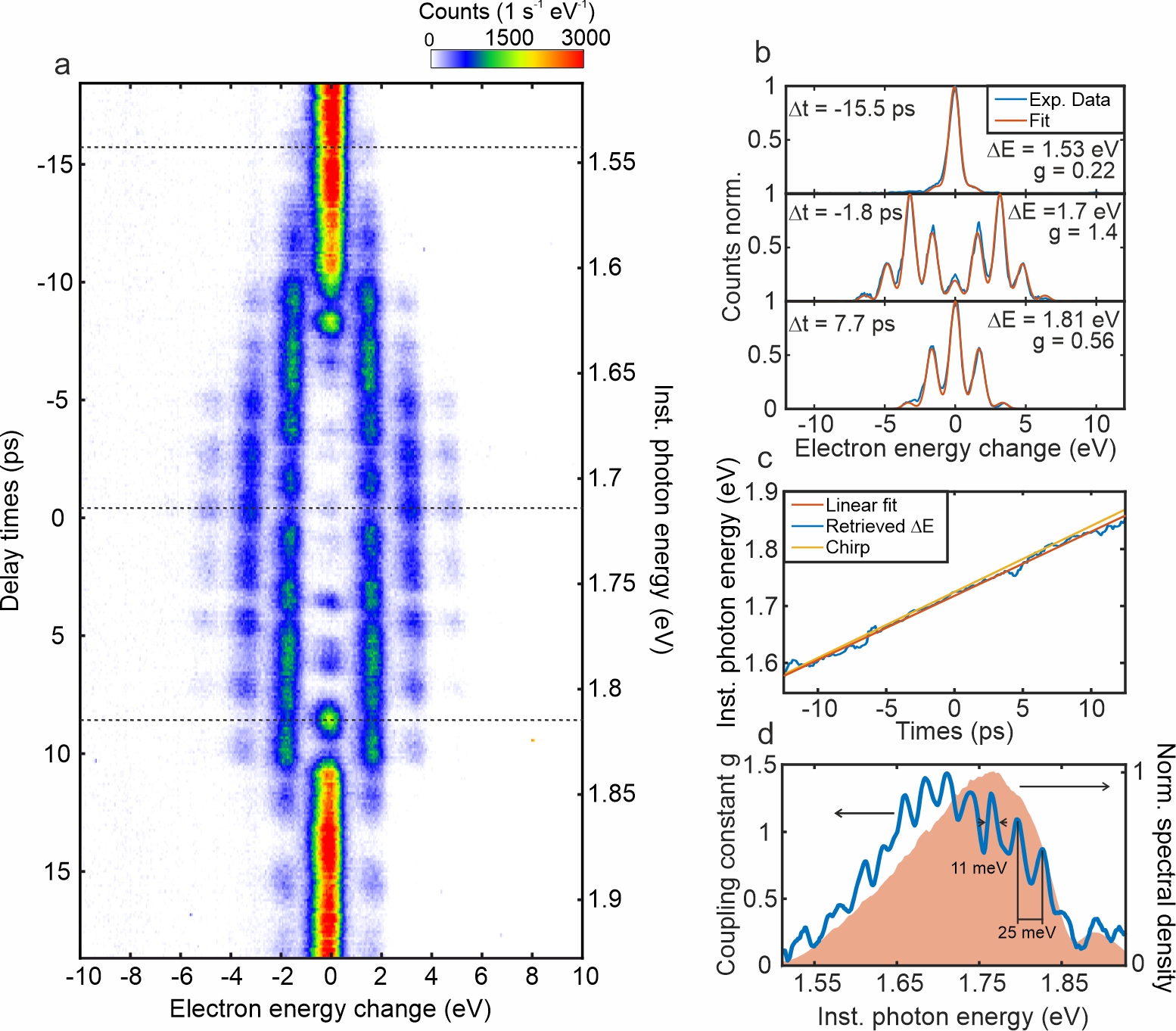}
	\caption{Spectrally-resolved electron-light coupling strength. \textbf{a} Electron spectogram accquired by changing the delay time between the electron and the optical pump pulse. A large number of electron energy sidebands corresponds to a higher coupling strength. \textbf{b} Exemplary electron energy spectra (blue) recorded at the delay times indicated by dotted lines in \textbf{a}. Spectra are fitted (red curves) using a model with the instantaneous photon energy and the coupling strength $g$ as free fitting parameters. \textbf{c} Sideband energy spacing $\Delta E$ retrieved from fitting the electron spectra for all delay times compared to the optical pulse chirp. A linear fit provides a relation between the delay time and the instantaneous photon energy as utilized in the right axis of the plot in \textbf{a}.  \textbf{d} Photon energy dependent coupling strength $g$ (blue) and spectral field distribution of the incident light pulse (shaded light orange). The coupling strength exhibits a photon energy dependent modulation of 25 meV and a shift of the coupling strength maximum relative to the peak of the optical spectrum. Features with a width down to 11 meV are resolved.}
	\label{figure2}
\end{figure}	

\section{Spectrally resolved electron-light coupling strength}

In Fig. 2a, we show the change in the electron energy spectra for different delay times. Considering the optical pulse chirp each delay time can be associated with an instantaneous photon energy (right axis in Fig. 2a). Thus, the change in the number of photon sidebands with delay is not only connected to the incident light spectrum but also the wavelength-dependent coupling strength which depends on details of the optical sample response. For example, the pronounced modulation in the number of sidebands observable over the complete delay range is not contained in the incident spectrum but attributed to spatial features in the light field close to the sample surface, as discussed below. In addition, for later delays the spacing between photon sidebands increases, in agreement with the electrons sampling the lower-wavelength part of the up-chirped pulse.

The electron-light coupling strength is quantified by the coupling parameter $g$, with which the relative amplitude of the $n$-th order sideband is given by $J_n(2|g|)^2$ where $J_n$ is the $n$-th order Bessel function of the first kind \cite{di2020free,garcia2021optical}. As shown in Fig. 2b for three exemplary delays, the electron energy spectra are well-described by a fit of two free parameters - the coupling constant $g$ and the instantaneous photon energy - without the requirement to include additional decoherence channels (see Supplement). Notably, clear signatures of Rabi oscillations are visible in the spectra, for example the destructive interference for the $n=1$ photon sideband occur for $\Delta t= -1.8$~ps (middle panel), further highlighting the spatial and temporal coherence of the electron-light interaction. The delay-dependent instantaneous photon energies as extracted from the spacing of photon sidebands (Fig. 2c) closely match the imprinted light chirp (yellow line). The overall spectral dependence of the extracted $g$ (Fig. 2d, blue line) follows to some extent the spectral amplitude of the incident light (orange shaded area). However, distinct differences between the light spectrum and interaction strength are visible. In particular, the coupling strength seems to exhibit a periodic modulation with a periodicity of about 25~meV. For precide absolute calibration of the energy axis, a spectral dip at 1.86 eV is introduced by an optical element placed in the pump beam, resulting in shift of 0.113~eV to match the corresponding drop in the coupling strength. 

\begin{figure}[H]
	\includegraphics[scale=1]{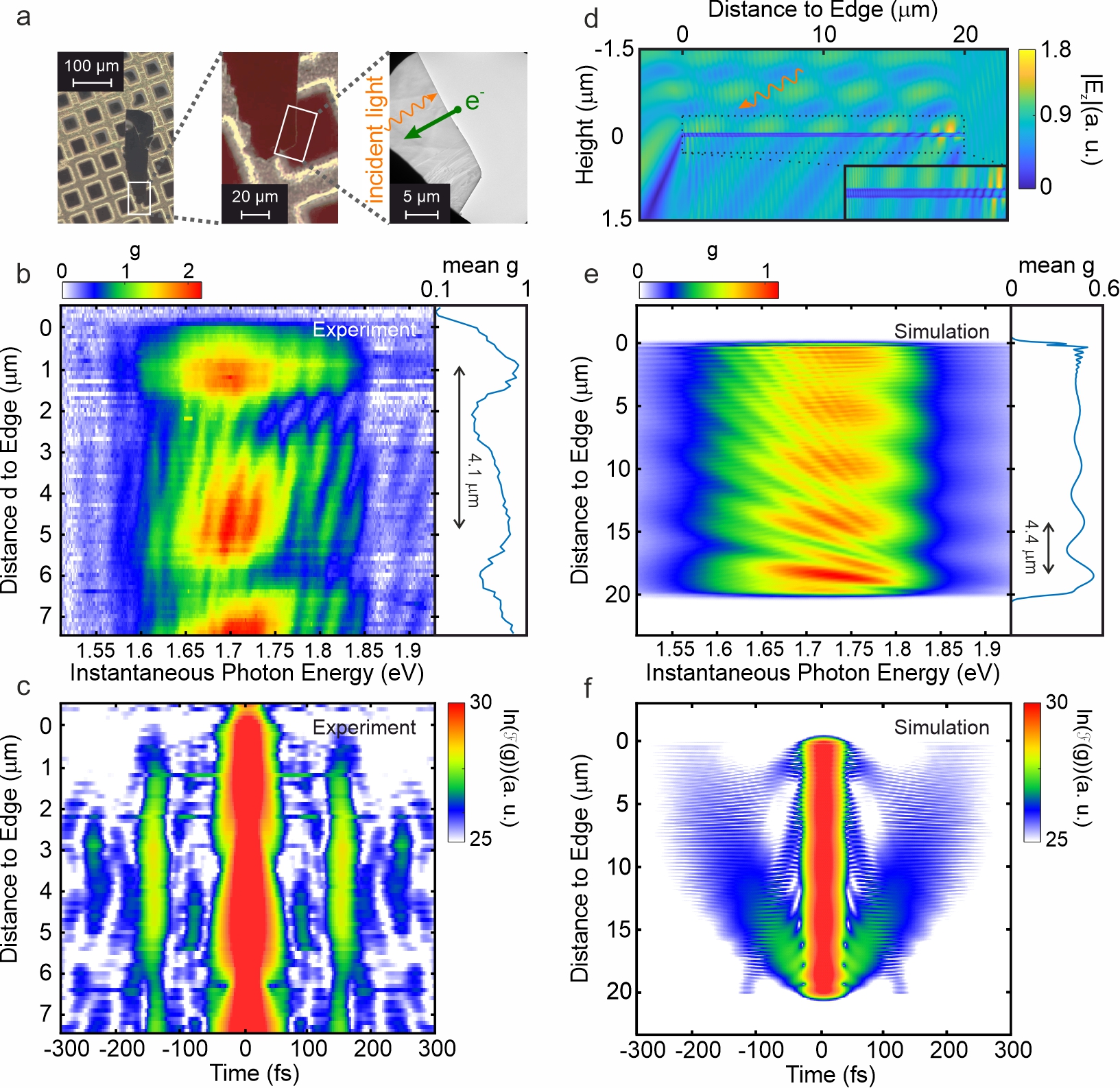}
	\caption{Spatial-dependence of electron-light coupling strength and numerical simulations. \textbf{a} Left/Center: Light microscopy images of the MoS$_2$ flake placed on a gold grid. Right: TEM image of the edge of the MoS$_2$ flake at which 
	the experiments are performed. The electron beam (spot size 10~nm) is moved along the scan direction (green arrow) perpendicular to the flakes edge and every position a spectrogram is recorded. The projected direction of the pump light is indicated by a wave orange arrow. \textbf{b} Photon-energy-resolved coupling strength $g$ recorded at varying distances from the MoS$_2$ flake edge, exhibiting a photon-energy dependent modulation as in Fig. 2a with a spatially dependent phase. The spectrally integrated coupling strength (right panel) is governed by an interference-related modulation of the near-field strength. \textbf{c} Fourier transform of the map in \textbf{b} along the energy direction. \textbf{d} Numerically calculated absolute value of the electric field component along z-direction (electron beam direction) close to a MoS$_2$ stripe (width: 20~µm, thickness: 70~nm) at a photon energy of 1.77 eV. \textbf{e} Spatio-temporally resolved coupling strength for the calculated field distribution in \textbf{d} and its spectral integral (right panel) showing a comparable spatial modulation to the experimental data. \textbf{f} Fourier transform of \textbf{e} along the photon energy direction.}
	\label{figure3}
\end{figure}

\section{Coupling strength at MoS$_2$ thin film edge}

To further analyze the spectrally periodic modulations in the coupling strength, we performed a one-dimensional spatial scan of the electron spot position in a direction perpendicular to the flake edge (8~µm scan range, 100~nm steps), as shown in Fig. 3a. The flake was positioned so that the pump beam polarization is perpendicular to the edge. At each position,  electron energy spectra for delay times between -18.4 and +18.4~ps (123~fs step size) are recorded. The photon-energy dependent coupling constant extracted from these spectrograms at each position is plotted in Fig. 3b. Notably, within this map spectral features with a FWHM down to 11 meV are discernible (see Fig. 2d ).
The photon-energy averaged coupling constants (right panel Fig. 3b) exhibits pronounced spatial variations with a periodicity of about 4.1~µm. The first maximum in the coupling strength is located at a distance of around 600 nm from the edge of the flake. In addition to these spatial variations, at each probing position the full distance-energy-resolved coupling maps show the distinct spectral modulation with a period of about 25~meV, as already observed in Fig. 2d but with the phase of the modulation shifting by 2.4~rad/µm (see Supplement). A Fourier transform of the map along the energy axis is depicted in Fig. 3c. Here, the 25~meV spectral oscillation corresponds to signals occurring at time periods of about 165~fs. For edge distances between 2 and 6~µm, the time difference does not show any spatial dependence. 

For a further understanding of spatial and spectral PINEM characteristics at the MoS$_2$ edge, we simulated the electric field distribution along the electron direction ($E_z$) for monochromatic incident plane waves of different photon energies (see Methods). The absolute value of the optical field distribution near the edge is exemplary depicted in Figure 3d for a photon energy of 1.77~eV and an incident wave vector along the orange wavy arrow. To approximate the experimental sample geometry, we consider a stripe-like MoS$_2$ flake with a width of 20~µm, a thickness of 70~nm and an infinite extension along the edge (quasi 2D geometry). Similar to the optical field incident on an infinite MoS$_2$ surface, a modulation of the field intensity along the surface normal is formed above the flake with a periodicity of 640 nm due to the interference between the incident and reflected wave components. An additional modulation in the direction perpendicular to the edge is observable above and below the sample due an interference between these two wave components and scattered cylindrical waves originating from each of the two edges. Due to different directions of both cylindrical waves, the components launched at $d=0$~µm result in an interference pattern with periodicity $L\approx 380$~nm ($L=\lambda/(1+\sin(\alpha))$). In contrast, the scattered wave started at $d=20$~µm yields a periodicity of $L\approx4.3$~µm ($L=\lambda/(1-\sin(\alpha))$). Furthermore, close to the sample surface, the interference of evanescent wave components of a guided mode with the external wave field is visible. This part of the interference pattern sensitively depends on the flake thickness (see Supplement) due to its impact on the mode index and the number of guided modes.

From the simulated optical fields, we retrieved the expected electron light coupling strength $g$ (see Methods) for different probing positions $d$ and illuminating photon energies. The resulting spatio-spectral map is plotted in Fig. 3e. We note that the map features largely resemble the field distribution at the flake surfaces, since $g$ is connected to spatial Fourier components outside the light cone and therefore strongly depends on discontinuities in the field. 
Comparing the simulation with experimental results (Fig. 3b), it is apparent that the order-of-magnitude of the electron-light coupling strength is well reproduced. In addition, the spatial modulation of $g$ with a periodicity of $L=4.4$~µm  (caused by scattering at the edge located at $d=20$~µm) is also found in the experiments. Curiously, the expected modulation with smaller periodicity is not observed experimentally, which could be related to the rather large experimental probing step size of 100 nm. Furthermore the pronounced spectral dependency of $g$ found both experimentally and theoretically shows distinct differences between Fig. 3b and e. Notably, the spectral fringes are tilted in opposite directions and spatial dependence of the spectral periods (Fig. 3c and f) is different in both cases. Several aspects are likely responsible for this discrepancy. Firstly, the sample geometry (Fig. 3a) is far from a perfect strip-like flake (Fig 3a middle panel) with a number of additional edges on the flakes side opposing the scan region, an additional close-by edge at a distant of around 10 µm from the scan region with an angle of 60$^{\circ}$ to the scan edge and a supporting gold grid bar which are expected to result in additional scattering sources. This would be further supported by the rather high fringe visibility in the (above bandgap) photon energy range from 1.77 eV to 1.82 eV for which guided Fabry-Perot modes would be strongly damped. Lastly, given the excitation strength we estimate an optically generated exciton density of up to 0.5 nm$^{-3}$ which should result in pronounced exciton-exciton interactions, as observed in ultrafast optical spectroscopy \cite{sun2014observation,chernikov2015population}, offering the exciting possibility of imaging dense exciton systems in ultrafast electron microscopy.  

\section{Conclusion}

In summary, we demonstrated the photon-energy-resolved spatial mapping of the electron-light coupling strength at an MoS$_2$ thin film edge by using broadband, strongly chirped light pulses. The energy spacing between the electron energy sidebands show a linear dependence on the temporal delay between the electron pulse and the chirped light pulse closely following the light pulses instantaneous frequency. By comparing the spectral dependency of the coupling strength to the light pulses spectral field distribution we found pronounced variations accross the spectrum due to the optical response of the thin film. This response was further analyzed performing a one-dimensional spatial scan with the electron beam over the MoS$_2$ thin film edge revealing both spectral and spatial modulations of the electron-light coupling. To back up our experimental findings we numerically simulated the electric field at a MoS$_2$ thin film and retrieved the expected electron-light coupling strength. The modulations in the coupling strength were traced back to interferences of the incident and reflected light waves with the waves scattered at the thin films edges or optical modes guided within the thin film. Our findings show that well-known PINEM effect can be utilized to map the spatially- and spectrally-resolved optical response of a semiconductor in the below- and near-bandgap region. As a next step, these investigations will be extended to semiconductors with intrinsic inhomogeneities, such as TMDC moiré bilayers, semiconductor heterostructures and atomic defects to gauge the fundamental limits of this technique in terms of spatial resolution and the retrieval of a local dielectric function.  

\section{Acknowledgment}
We acknowledge financial support by the DFG within the priority program 1840 "Quantum Dynamics in Tailored Intense Fields", funding by the Volkswagen Foundation as part of the Lichtenberg professorship "Ultrafast nanoscale dynamics probed by time-resolved electron imaging" and funding by the Free State of Bavaria through the Lighthouse project "Free-electron states as ultrafast probes for qubit dynamics in solid-state platforms" within the Munich Quantum Valley initiative. In addition, we thank Thomas Quenzel, Petra Groß and Christoph Lienau for providing information and help for setting up the NOPA system. 
	
\section{Author Contributions}
N.M. performed the experiments with the help of S.e.K., G.V. and C.R.. Numerical calculations were performed by N.M. with contributions from S.e.K.. The sample was prepared by L.H.. N.M. and S.S. wrote the manuscript together. S.S. supervised and conceived the project.

\section{Methods}
\subsection{Experimental setup}
The experiment was conducted using a 200-kV ultrafast transmission electron microscope (UTEM) based on a commercial TEM (JEOL JEM F200) which we modified for allowing femtosecond optical-pump/electron-probe capabilities. For achieving bright electron pulses, we developed a laser-driven cold-field emitter source together with JEOL Ltd., allowing for side-entry optical excitation \cite{schroederinprep}. Ultrashort electron pulses are generated by single-photon photoemission from the tip of the cold field emitter gun (CFEG) using femtosecond ultraviolet pulses (wavelength: 342~nm, repetition rate: 400~kHz, pulse width: 167~fs) and an extraction field at the tip of around 1.9 GV/m. The generated electron bunches contain 1 electron per pulse on average with a pulse duration of 200~fs. For driving the optical near-field at the sample, we used the output of a home-build noncollinear optical parametric amplifier (NOPA) generating broadband light pulses (spectral range $@$-1db: 650 to 800~nm). These pulses are elongated in time by propagation through two 10~cm long cylinders of strongly dispersive dense flint glass (H-ZF52, group velocity dispersion (GVD) of 271.51~fs$^2$/mm @ 800~nm), resulting in an overall time domain broadening of the up-chirped optical pulse of up to 25~ps. In this limit the temporal pulse form almost equals the shape of its spectral field distribution. 
 
Electron pulses pass through the sample parallel to its surface normal and the light pulse impinges on the sample with an angle of about $57^{\circ}$ relative to the surface normal (estimated convergence half angle of 2.5$^{\circ}$). All reported experiments were performed with an optical power at the sample of 3.5 mW. Electron energy spectra are recorded using an electron energy filter (CEOS CEFID) and a sensitive CMOS camera (Tietz TemCam XF416). Spatial scans are recorded by shifting the focused electron beam (Spot 3, Alpha 5, 200-µm condenser aperture, 10-nm electron focal spot size) during consecutive temporal scans using the electron microscopes beam deflector coils. In order to ensure a proper in-coupling into the spectrometer over the complete spatial scan, a electron de-scanning routine was implemented below the sample. The laser beam was strongly defocused on the sample surface (estimated spot size of 50~µm) to ensure a homogeneous field distribution over the spatial scan region and for minimizing chromatic aberrations. \\ \\

\subsection{Numerical calculations}	
Numerical simulations were performed using the Ansys Lumerical software suite. To calculate the electric field distributions around the MoS$_2$ thin film (permittivity tensors taken from Ref.~\cite{munkhbat2022optical}) we used Lumerical's finite-difference time-domain (FDTD) solver in a 120 x 11 µm$^2$ 2D frame perpendicular to the films surface (assuming quasi infinite film extension along the edge). As boundary conditions, a perfectly matched layer was applied. For illumination, a broadband plane wave (Gaussian spectral shape, center wavelength: 725~nm, full-width-at-half-maximum (intensity): 150~nm) was utilized incident under an angle of 57$^{\circ}$. The wavelength resolved electric field distributions were retrieved via a frequency-domain monitor and the electric field strength was then calculated from the laser pulse parameters in the experiment. The electron light coupling strength $g$ was retrieved from the field distribution as a spatial Fourier transform along the electron beam direction (parallel to surface normal, z-direction) according to \cite{park2010photon}: 
	\begin{equation}
		g=\frac{e}{\hbar\omega}\int E_z(z) e^{-i\Delta k z}dz,
	\end{equation}
	with $\Delta k=\omega/v_e$ being the momentum change of a electron moving with velocity $v_e$ when a photon of frequency $\omega$ is absorbed or emitted. 
	
\subsection{Supporting information}
Supporting information: Details on the NOPA-setup, details on fitting procedure for electron energy spectra, coupling strength scan on SiN, phase shift of oscillation in coupling strength, simulation of coupling strength for different flake sizes, electric field distribution around the thin film (PDF)

	\bibliography{bibliography.bib}

\end{document}


\section{Non-collinear optical parametric amplifier}
\begin{figure}[H]
	\includegraphics[scale=2]{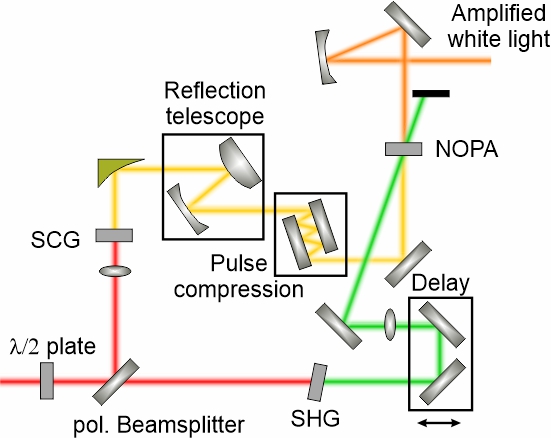}
	\caption{Layout of non-collinear optical parametric amplifier (NOPA) used to generate broadband optical pulses. Description in the text.}
	\label{nopa}
\end{figure}

For generating broadband optical pump pulses, we used a home-build non-collinear optical parametric amplifier (NOPA). The layout is depicted in Fig. \ref{nopa}. The incoming fundamental beam (1028 nm, 17 W @ 400 kHz) is split using a polarizing beamsplitter and a $\lambda/2$-plate to tune the fraction of the beam that is reflected into the signal arm and the fraction that is transmitted into the pump arm of the NOPA. The fundamental beam in the signal arm was adjusted to hold a pulse energy of 1 $\mu$J and is focused using a 100 mm lens into a 4 mm thick yttrium aluminium garnet (YAG) crystal for super-continuum generation (SCG) providing a stable broadband white light (500-1100 nm). A parabolic gold mirror is used to collimate the generated white light. A slightly de-adjusted reflection telescope build from a convex and concave mirror is then used to focus the beam into a $\beta$ barium borate (BBO) crystal while passing through chirped-mirrors used for temporal compression of the generated white light. In the pump arm, a separate BBO crystal is used for second harmonic generation (SHG). The frequency doubled beam passes through a retro reflector placed on a delay stage and is focused using a 200 mm lens. The SHG pump beam is spatially and temporally overlapped with the white light pulse in the BBO crystal used for non-collinear parametric amplification (NOPA) and the angle between both beams is adjusted to provide broadband amplification of the white light.          

\section{Fitting electron energy spectra and retrieve instantaneous photon energy}
To retrieve the electron photon coupling strength $g$ from the measured electron energy spectra, we employ a number of Gaussian functions displaced by $n$ times the instantaneous photon energy with an amplitude determined by the coupling strength:
\begin{equation}
	\sum_{n=-10}^{10}\frac{J_n(2|g|)^2}{\sigma \sqrt{2\pi}}\exp\left(-\frac{(E-E_0-n\Delta E)^2}{2\sigma^2}\right),
\end{equation}
where $n$ is the sideband photon order, $J_n$ is the n-th Bessel function of the first kind, $\sigma$ is the variance of the gaussian function, $E_0$ is the energy position of the zero loss peak and $\Delta E$ the instantaneous photon energy. The variance $\sigma$ and $E_0$ where determined beforehand by fitting spectra in which only the zero order peak ($n=0$) is present, so that $g$ and $\Delta E$ are the only free fit parameters for the light modulated spectra. The retrieved $\Delta E$ increases linearly with the time delay between optical pump pulse and electron pulse with a gradient close to the one expected from the applied chirp. To account for a slight miscalibration of the energy filter (about 0.113 eV), we used the spectral dip at 1.86 eV introduced by an optical element placed in the beam path to align the instantaneous photon energy axis leading to the result in Fig 2c of the main text.

\section{Spatial Scan on silicon nitride}
\begin{figure}[H]
	\includegraphics[scale=1]{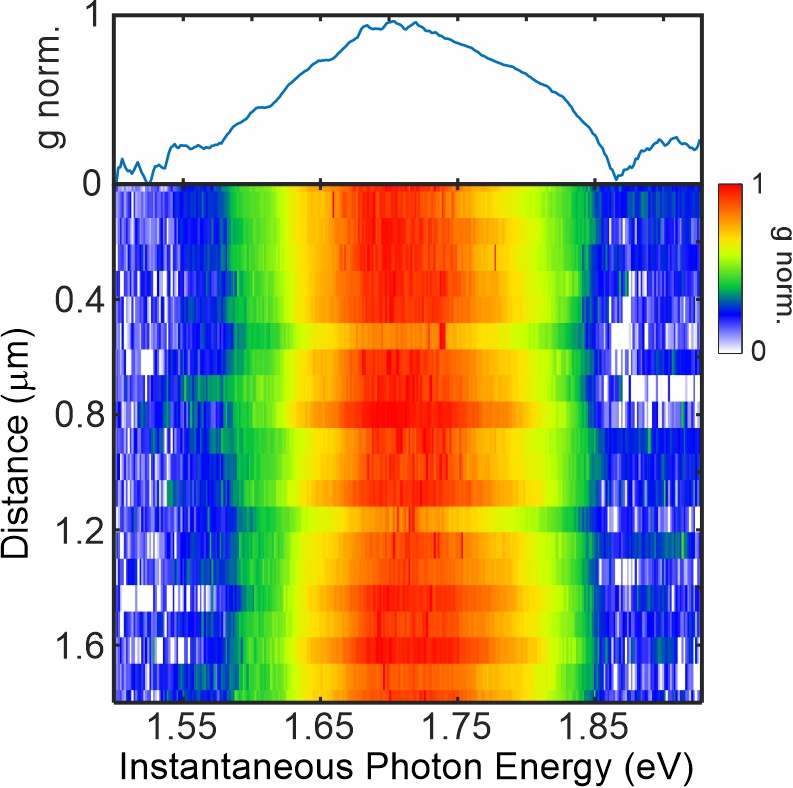}
	\caption{Spatial and spectrally resolved electron light coupling strength on a homogeneous silicon membrane. The scan shows no spatial and spectral dependencies except for the shape of the optical pump pulses spectral field distribution. The top panel shows the normalized sum over all spatial scan positions.}
	\label{SiN_figure}
\end{figure}

Additionally to the scan performed on the MoS$_2$ flake presented in the main text, we performed a spatial scan on a large homogeneous 200-nm-thick silicon nitride membrane (Fig. \ref{SiN_figure}) for comparison. The scan shows no spatial dependency and the coupling strength exclusively follows the pump pulse' spectral field distribution as expected. By this scan we also ensure that none of the features discussed for the MoS$_2$ measurement in the main text are a result of artifacts like for example optical interferences in the laser focus.      

\section{Phase of fast oscillation}

\begin{figure}[H]
	\includegraphics[scale=1]{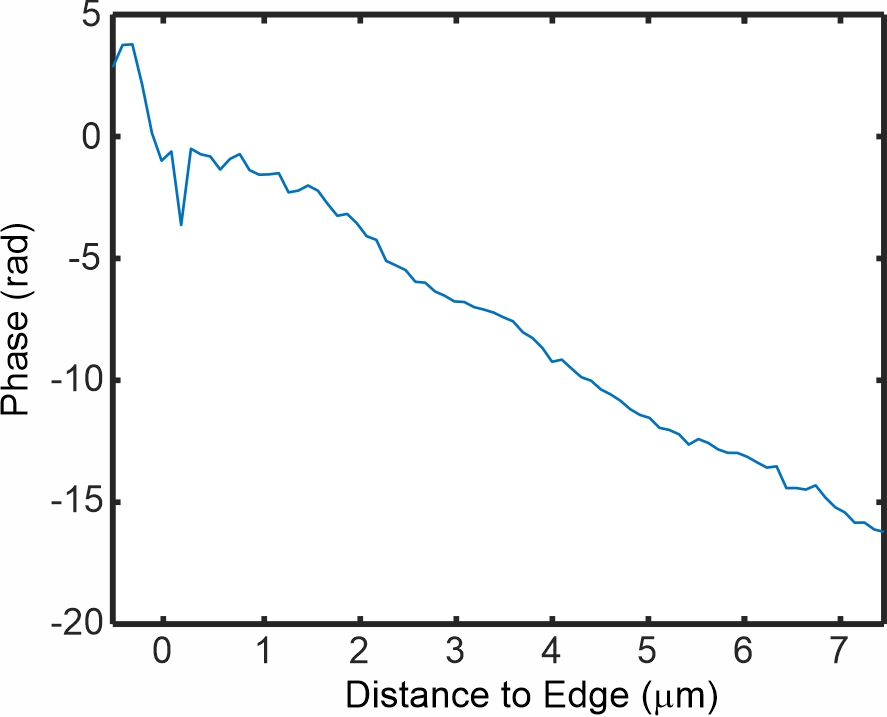}
	\caption{Phase of the Fourier component around 165~fs from Figure 3c of the main text.}
	\label{phase}
\end{figure}

Figure \ref{phase} shows the phase of the Fourier component around 165~fs from Figure 3c of the main text over the position of the electron beam on the sample surface. 

\section{Simulation of coupling strength for different flake sizes}

\begin{figure}[H]
	\includegraphics[scale=1]{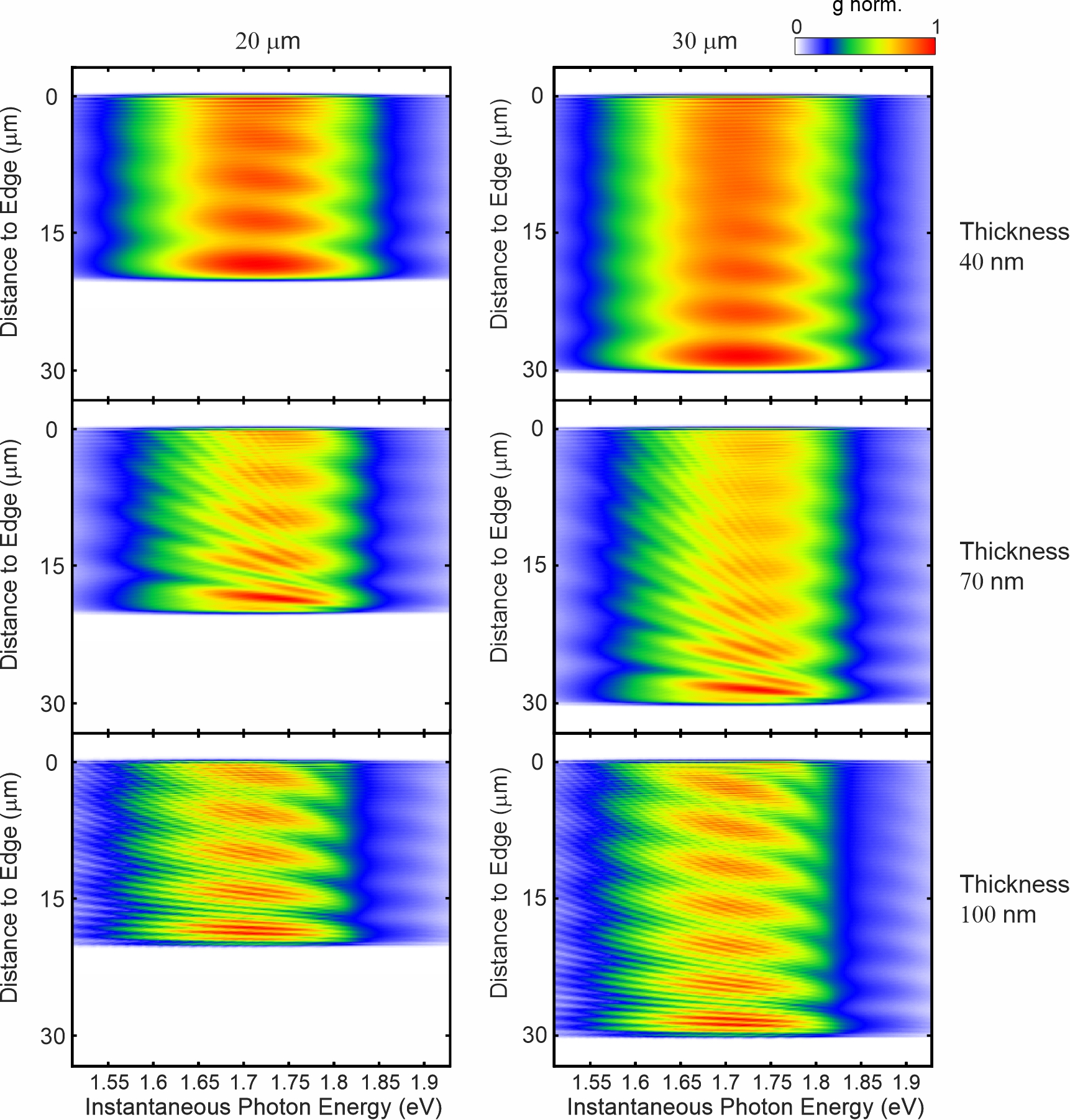}
	\caption{Simulated electron light coupling strength $g$ on MoS$_2$ depicted for increasing thin film thickness (top to down 40~nm, 70~nm, 100~nm) and to different flake sizes (20~$\mu$m and 30~$\mu$m).}
	\label{sim}
\end{figure}

For better comparability and to gain more insight in the overall behavior of the electron light coupling strength at the MoS$_2$ edge, we simulated $g$ for different thin film thicknesses and flake sizes some of them exemplary shown in Figure \ref{sim}.   
The coupling strength in the case of a 40~nm thick flake is governed solely by the interference pattern resulting from the interference between the cylindrical waves scattered at the flakes edges and the incident and reflected field. Due to the small thickness of the thin film no guided modes exist within the film. In contrast, for a 100~nm flake the influence of the guided mode is visible again due to inference with the incident wave but with a shorter modulation period along the flake compared to 70~nm. On the other side the overall size of the flake does not change the periodicity of the different modulations.        	

\section{Electric field distribution around the thin film}
\begin{figure}[H]
	\includegraphics[scale=1]{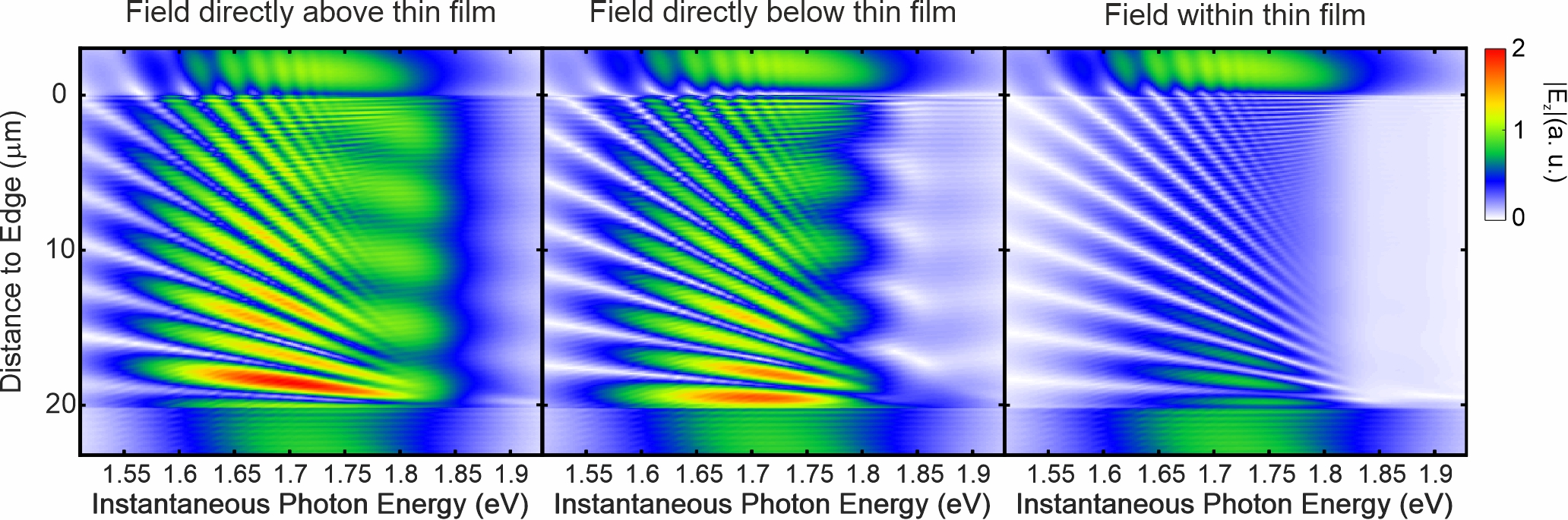}
	\caption{From left to right, a cross section of the electric field directly above, below and within the thin film are depicted.}
	\label{filed_above_in_below}
\end{figure}

The electric field directly above and below the sample as well as the field within the sample provide the necessary momentum broadening to couple to the electron beam due to its abrupt change at the interfaces. Therefore in Figure \ref{filed_above_in_below} cross sections of the respective fields are shown. While the field above and below the flake show comparable features to the coupling strength map presented in figure 3 of the main text (with respective phase shifts) the field distribution within the sample is mainly dominated by interference pattern caused by the guided mode.   
